\documentclass[prb,twocolumn,amsmath,amssymb,aps,floatfix,superscriptaddress]{revtex4-2}
\usepackage[version=3]{mhchem} 

\usepackage{amssymb}
\usepackage{color}
\usepackage{siunitx}
\usepackage{amsmath,bm}

\DeclareMathAlphabet{\mathpzc}{OT1}{pzc}{m}{it}

\usepackage{graphicx}
\usepackage{dcolumn}
\usepackage{bm}

\begin{document}

\author{Rai M. Menezes}%
\affiliation{Departamento de F\'isica, Universidade Federal de Pernambuco, Cidade Universit\'aria, 50670-901, Recife-PE, Brazil}
\affiliation{NANOlab Center of Excellence \& Department of Physics, University of Antwerp Groenenborgerlaan 171, B-2020 Antwerp, Belgium}
\author{Milorad V. Milo\v{s}evi\'c}%
\email{milorad.milosevic@uantwerpen.be}
\affiliation{NANOlab Center of Excellence \& Department of Physics, University of Antwerp Groenenborgerlaan 171, B-2020 Antwerp, Belgium}

\title[An \textsf{achemso} demo]
  {Skyrmion blinking from the conical phase}

\keywords{spintronics}

\begin{abstract}
While the transition between skyrmionic and non-topological states has been widely explored as a bit operation for information transport and storage in spintronic devices, the ultrafast dynamics of such transitions remains challenging to observe and understand. Here, we utilize spin-dynamics simulations and harmonic transition state theory (HTST) to provide an in-depth analysis of the nucleation of skyrmionic states in helimagnets. We reveal a persistent blinking (creation-annihilation) phenomenon of these topological states under specific conditions near the phase boundary between skyrmion and conical states. Through a minimum-energy path analysis, we elucidate that this blinking behavior is favored by the formation of chiral bobber (CB) surface states and that the collapse of CBs differs from that of skyrmions in thin films due to their different oscillation modes. We further employ HTST to estimate the typical blinking time as a function of the applied magnetic field and temperature. Finally, we illustrate the practical use of skyrmion blinking for controlled probabilistic computing, exemplified by a skyrmion-based random-number generator.
\end{abstract}

\maketitle

\section{Introduction}

The interest in non-traditional computing architectures and in-memory systems is growing due to their ability to tackle problems that conventional computers struggle with. Spintronic technologies, such as stochastic magnetic tunnel junctions~\cite{grollier2020neuromorphic} and nano-oscillators~\cite{bohnert2023weighted}, are promising for their low energy consumption and high-speed computing capabilities.  Particularly, magnetic skyrmions -- topologically protected quasiparticles -- show potential for unconventional computing applications~\cite{back20202020,li2021magnetic,grollier2020neuromorphic}, offering the stability, compacted size, and facilitated manipulation with low energy consumption as needed for modern applications. 
The intense research effort over the years led to a range of skyrmionic technology, such as proposals for magnetic memory devices and logic operators~\cite{luo2021skyrmion}, skyrmion-based Qubits~\cite{psaroudaki2021skyrmion,petrovic2021skyrmion}, reservoir computing~\cite{prychynenko2018magnetic,yokouchi2022pattern}, and many others. Recent works also suggested that thermally induced skyrmion dynamics can be used for probabilistic computing devices~\cite{pinna2018skyrmion,zazvorka2019thermal,wang2022single}.
In most cases, realizing such applications necessitates the ability to intentionally generate and eliminate individual skyrmions, as well as manipulate their spatial and temporal positioning~\cite{reichhardt2022statics}.

It is well known that magnetic skyrmions can emerge from the conical phase in chiral helimagnets such as MnSi and FeGe as the system is excited by the appropriate applied magnetic field and temperature~\cite{muhlbauer2009skyrmion,leishman2020topological}. Understanding the nucleation mechanism of magnetic skyrmions in such materials can potentially grant the capability to control their stability~\cite{stosic2017paths,muckel2021experimental,paul2020role} and orchestrate their spatial and temporal arrangement, crucial for various applications. Recent works aiming to achieve such understanding have shown the temporal evolution of skyrmion nucleation from the conical phase in thin plates of Co$_8$Zn$_{10}$Mn$_2$~\cite{Kim2020mechanism,kim2021kinetics}. Among other interesting features, such as clustering, the process of skyrmion-lattice (SkL) formation is characterized by a combination of skyrmion nucleations and collapses. We suggest that this blinking behavior (creation-destruction process) is favored by the local stability of an intermediate state between the conical and skyrmion phases (observed in Ref.~\citenum{Kim2020mechanism} with about $60\%$ of the contrast of a fully formed skyrmion, dubbed a skyrmion embryo), the so-called chiral bobber (CB)~\cite{zheng2018experimental} — a topologically protected swirl of the magnetization localized at the material's surface — which breaks the nucleation process into stages with a lower energy cost compared to homogeneous skyrmion nucleation~\cite{rybakov2015new}. Once at this intermediate state, the energy barrier for the system to transition to the skyrmion phase can be similar to the barrier to return to the conical phase. This could lead to sequential instances of nucleation and collapse of CBs (skyrmion embryos) during the formation of the skyrmion lattice, as seen, for example, in Lorentz transmission electron microscopy (LTEM) images in Ref.~\citenum{Kim2020mechanism}. In fact, a detailed description of such rich dynamics remains elusive to date.

In this work, we provide an in-depth analysis of skyrmion blinking behavior and demonstrate that the skyrmion system can be set to uniform blinking when gauged by the appropriate applied field and temperature. We show that this condition is favored when the system is close to the phase boundary between the SkL and conical states, specifically on the conical side of the phase boundary, where CBs are metastable. Additionally, we calculate the oscillation modes of the CB states to demonstrate that their collapse differs from that of skyrmions, as a different number of zero modes of oscillation are associated with the two states. We then use transition state theory to estimate the typical blinking time of CBs as a function of the applied field and temperature, and exemplify the use of skyrmion blinking for controlled probabilistic computing, such as in a random-number generator.

The paper is organized as follows. In Sec.~\ref{secII}, we provide the
analytic considerations and describe the spin model used to simulate the magnetic system (Sec.~\ref{secIIA}), as well as the geodesic nudged elastic band method used to calculate minimum energy paths along the magnetic phase transitions (Sec.~\ref{GNEB}) and the harmonic transition state theory considered to estimate the state lifetime (Sec.~\ref{HTST}). In Sec.~\ref{sec.III}, we report the phase diagrams of the considered magnetic films with different thicknesses (Sec.~\ref{sec.IIIA}) to then investigate the nucleation mechanism of skyrmions from the conical phase, as well as the dependence of their activation energies on the applied field and film thickness (Sec.~\ref{secIIIB}). In Sec.~\ref{sec.IIIC}, we detail the formation of fully formed skyrmions, after nucleation of CBs. Sec.~\ref{sec.IV} is devoted to the skyrmion blinking dynamics, investigated by spin dynamics simulations (Sec.~\ref{sec.IVA}) and transition state theory (Sec.~\ref{sec.IVB}). Finally, in Sec.~\ref{secV}, we exemplify the use of skyrmion blinking for controlled probabilistic computing, and demonstrate the basic functionalities of a skyrmion-based random-number generator. Our results
are summarized in Sec.~\ref{Sec:Conclusion}. 

\section{Theoretical framework}\label{secII}

\subsection{Spin model}\label{secIIA} 

To investigate the formation of magnetic skyrmions in chiral helimagnets, we perform spin-dynamics simulations by employing the numerical package SPIRIT~\cite{muller2019spirit}. We make use of the extended Heisenberg Hamiltonian that describes the system of classical spins and can be written as
\begin{equation}
\begin{aligned}
  \mathcal{H}
    = & -\sum_{\langle i,j \rangle} \mathcal{J}_{ij}\textbf{n}_i \cdot \textbf{n}_j - \sum_{\langle i,j \rangle} \textbf{D}_{ij} \cdot (\textbf{n}_i \times \textbf{n}_j)\\
      & -\sum_i \mu\textbf{B} \cdot \textbf{n}_i,
    \label{HeisenbergHam}
\end{aligned}
\end{equation}
where $\bm{\mu}_i=\mu \textbf{n}_i$ is the magnetic moment of the $i^{\text{th}}$ spin site, with magnitude $\mu$ and orientation
$\textbf{n}_i$. 
Here, $\mathcal{J}_{ij}$ represents the exchange interaction, $\textbf{D}_{ij}$ is the Bloch-type Dzyaloshinskii-Moriya (DM) vector, $\textbf{B}$ is the applied magnetic field, and $\langle i,j \rangle$ denotes pairs of spin sites $i$ and $j$ accounting up to second nearest-neighbour (NN) sites. For the simulations, we adopt a spin-system parametrized by the first NN exchange interaction $\mathcal{J}_1 = J_\text{ex} = 1$~meV. The NN DM interaction is chosen as $D_1=\tan(2\pi/10)J_\text{ex}\approx0.727 J_\text{ex}$, which corresponds to a typical pitch length $L_D$ of ten lattice sites for a helix state at zero magnetic field. Next-nearest-neighbor interactions are chosen according to Ref.~\citenum{buhrandt2013skyrmion}, $\mathcal{J}_2 = J_\text{ex}/16$ and $D_2 = D_1/8$, which leads to a good representation of the magnetic phase diagram of helimagnets. The dynamics of the spin system is governed by the Landau-Lifshitz-Gilbert (LLG) equation
\begin{equation}
  \frac{\partial \textbf{n}_i}{\partial t}
    = -\frac{\gamma}{(1+\alpha^2)\mu} \left[\textbf{n}_i \times \textbf{B}_i^\text{eff} + \alpha\textbf{n}_i \times (\textbf{n}_i \times \textbf{B}_i^\text{eff}) \right],
\end{equation}
where $\gamma$ is the electron gyromagnetic ratio, $\alpha$ is the Gilbert damping parameter and $\textbf{B}_i^\text{eff} = -\partial \mathcal{H}/\partial\textbf{n}_i$ is the effective field. For the cases where it is applicable, temperature-dependent simulations are implemented by the introduction of a stochastic thermal field $\textbf{B}^\text{th}$, which is added as a contribution to the effective field acting on the localized spin-sites, i.e., $\textbf{B}_i^\text{eff}\rightarrow-\partial\mathcal{H}/\partial \textbf{n}_i+\textbf{B}_i^\text{th}$. The magnitude of the thermal field is obtained by the fluctuation-dissipation theorem, and is given by $\textbf{B}_i^\text{th}(T,t)=\bm{\eta}_i(t)\sqrt{2\mathcal{D}/\Delta t} =\bm{\eta}_i(t)\sqrt{\frac{2\alpha k_B T}{\gamma\mu\Delta t}}$, where $T$ is the temperature, $k_B$ is the Boltzmann constant, $\Delta t$ is the time-step considered in the simulation, and $\bm{\eta}_i(t)$ is a Gaussian white noise that represents the thermal fluctuations on each spin site $i$. The ensemble average and variance of the thermal field satisfies $\langle \textbf{B}_{i}^\text{th}(t)\rangle=0$ and $\langle\textbf{B}_{ia}^\text{th}(t)\textbf{B}_{jb}^\text{th}(t')\rangle=2\mathcal{D}\delta_{ij}\delta_{ab}\delta(t-t')$, respectively, where $a,b$ indicate the components of the vector $\textbf{B}_i^\text{th}$. The stochastic LLG equation provides equivalent results for the magnetic ground state as those obtained by Monte Carlo methods~\cite{muller2019spirit}.

\subsection{Minimum energy paths}\label{GNEB}

To characterize the dynamics of skyrmions during their nucleation from the conical phase, it is crucial to determine the nucleation mechanism and the energy barriers involved in the phase transition. To do so, we calculate the minimum energy paths (MEP) for skyrmion nucleation by making use of the geodesic nudged elastic band method (GNEB)~\cite{bessarab2015method} with the assistance of a climbing image (CI) method~\cite{henkelman2000improved}, both readily implemented in SPIRIT~\cite{muller2019spirit}. These methods enable an accurate determination of the highest-energy configuration, or saddle point, along the minimal energy path connecting the two states, from where the activation energy of the phase transition is determined.

In the GNEB method, we consider a path given by a discrete chain of $N_I$ magnetic configurations, called ``images" of the system, between two considered magnetic states. The initial guess of the path is then represented by the set of images $[\bm{\mathcal{M}}^1,...,\bm{\mathcal{M}}^{N_{I}}]$, where $\bm{\mathcal{M}}^\nu=(\textbf{n}_1^\nu,\textbf{n}_2^\nu,...,\textbf{n}_N^\nu)$ represents the magnetic configuration of the $\nu^{th}$ image of the system with $N$ spins. In our calculations, we generate these initial configurations by applying a homogeneous rotation of magnetic moments between the two magnetic states under consideration. In order to converge from the initial guess to the nearest MEP, the effective force at each image is calculated by the negative energy gradient $-\nabla \mathcal{H}^{\nu}$, where $\mathcal{H}^{\nu}$ is the energy of the $\nu^{th}$ magnetic configuration and $\nabla_i=\partial/\partial \textbf{n}_i$. The force component along the local tangent to the path is then substituted by an artificial spring force  between the images, which ensures uniform distribution of the images along the path, while the energy gradient forces orthogonal to the path tangents are applied, thus moving the images towards the minimum energy position. The first and last images of the chain are fixed and given by the local minima corresponding to the initial and final states.  In the CI method, the spring forces acting on the highest energy image are deactivated during the iterative optimization, and the energy gradient force is inverted to point along the path. This procedure makes the image to move uphill in the energy landscape along the path. After the CI-GNEB calculation has converged, the position of the CI coincides with the saddle point along the MEP and gives an accurate value of the activation energy, defined by the energy difference between the saddle point and the initial state.


\begin{figure*}[!t]
    \centering
    \includegraphics[width=\linewidth]{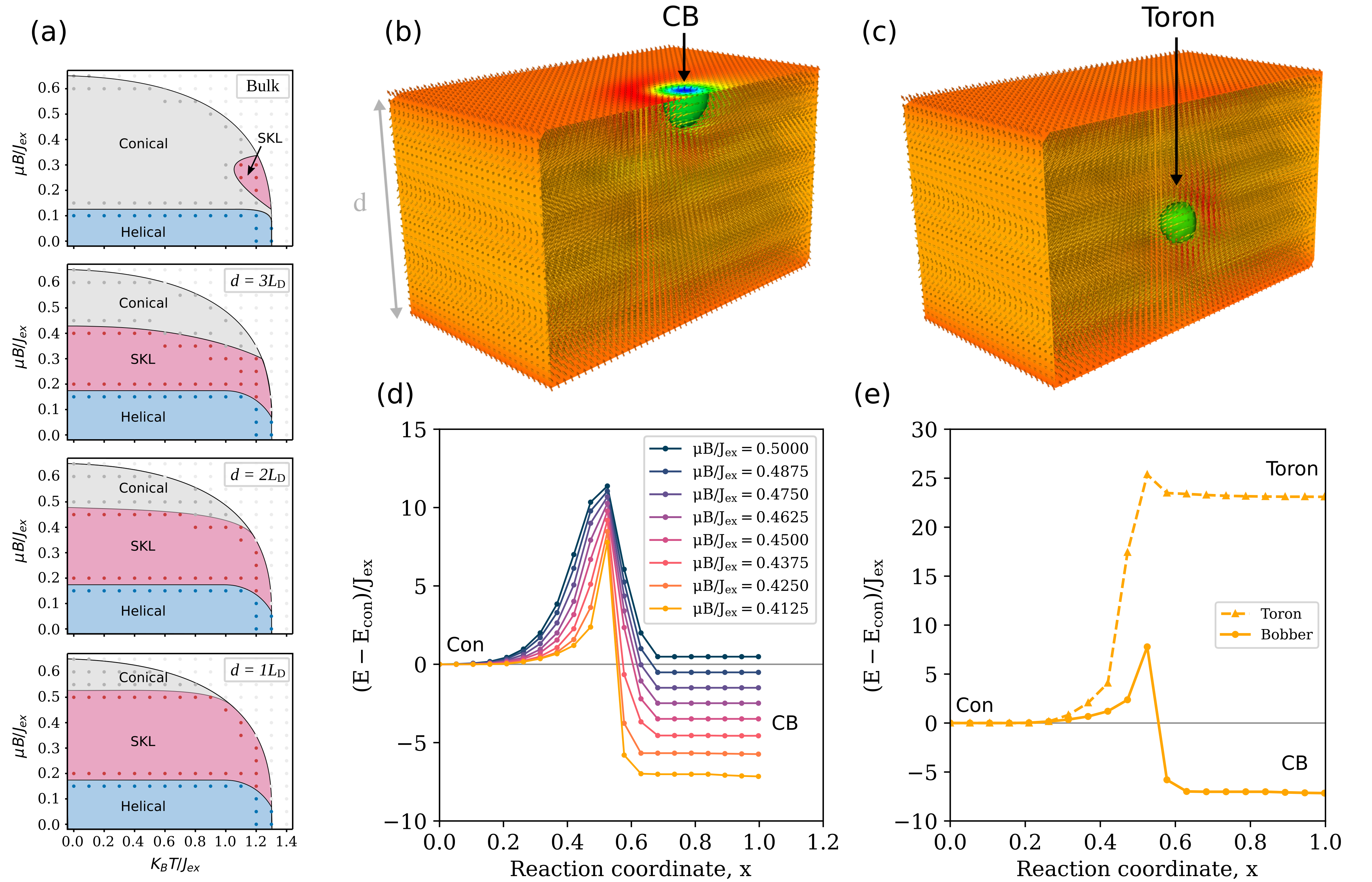}
    \caption{(a) Magnetic phase diagrams for films of different thickness, as obtained in the simulations. Here, the dots represent the lowest energy configurations obtained in the simulations, including the HL (blue), CON (gray), and SkL (red) phases. Paramagnetic and field-polarized phases are shown as light gray dots. (b-c) Exemplified chiral bobber (b) and toron (c) phases, as obtained in the spin-resolved simulations. (d) Minimum energy paths for CB nucleation, for different values of the applied magnetic field, for film of thickness $d=2L_D$. Here, $x=0$ corresponds to the conical state and $x=1$ corresponds to the CB state. (e) Comparison of the minimum energy paths for CB and toron nucleation, for film thickness $d=2L_D$ and applied field $\mu B/J_\text{ex}=0.4125$. }
    \label{fig1}
\end{figure*}

\subsection{Harmonic transition state theory}\label{HTST}

Once the energy barriers involved in the skyrmion nucleation and annihilation processes are known, the rate of such phase transitions under thermal fluctuations can be estimated by making use of harmonic transition state theory (HTST)~\cite{bessarab2012harmonic}, which assumes a Boltzmann distribution within the region of configuration space that corresponds to the initial state, i.e., before the system escapes (overcomes the energy barrier) due to thermal fluctuations. This assumption is valid when escape events are rare compared to the timescale of magnetization dynamics. The transition rate, $\Gamma$, is then given by an Arrhenius-type law, with exponential dependence on the inverse temperature $T$ and the energy barrier $\Delta E$, and can be written as
\begin{equation}
    \Gamma=\Gamma_0e^{-\Delta E/k_BT}.
    \label{eq.HTST}
\end{equation}
The pre-exponential factor, $\Gamma_0$, also known as the \textit{attempt frequency}, is 
dictated by the energy surface's curvature -- the Hessian matrix -- at both the saddle point and the initial state minimum, and is given by~\cite{bessarab2012harmonic,muller2019spirit}
\begin{equation}
    \Gamma_0=\frac{\gamma}{2\pi}v\sqrt{\frac{\prod_i^\prime \lambda_i^\text{min}}{\prod_i^\prime´\lambda_i^\text{SP}}},
    \label{eq.prefactor1}
\end{equation}
with
\begin{equation}
    v=\frac{V^\text{SP}}{V^\text{min}}\sqrt{2\pi k_B T}^{(N_0^\text{min}-N_0^\text{SP})}\sqrt{\sum_i{}^{\prime} \frac{a_i^2}{\lambda_i^\text{SP}}}.
    \label{eq.prefactor2}
\end{equation}
Here, $\lambda_i$ are the eigenvalues of the Hessian matrix, calculated at the initial state (min) and saddle point (SP).  $N_0$ is the number of zero modes -- modes with $\lambda=0$ -- and $V$ denotes the phase space volume associated with the zero modes. $a_i$ are expansion coefficients of the velocity along the unstable mode of the saddle point~\cite{bessarab2012harmonic}. The primes next to products and sum indicate that the terms associated with the unstable ($\lambda<0$) and zero modes are omitted from the calculation. The transition time (or lifetime) of the process is finally estimated as $\tau=1/\Gamma$.

\section{Skyrmion nucleation from the conical phase}\label{sec.III}

\subsection{Magnetic phase diagrams}\label{sec.IIIA}

Before investigating the nucleation mechanisms of the skyrmion phase, we first calculate the magnetic ground states of the considered spin system. For the simulations, we consider a sample of size $2L_D\times L_D\sqrt{3}\times d$, where we explore three different thickness values for the magnetic film, denoted as $d=L_D$, $d=2 L_D$, and $d= 3 L_D$, with $L_D$ representing the pitch length, discretized into 10 spin sites. In all three cases, periodic boundary conditions are set along the film plane, while open boundary conditions are applied across the film thickness. Additionally, we simulate the bulk system by considering $d= 3 L_D$  and periodic boundary conditions across the film thickness. To construct the equilibrium phase diagrams of the magnetic system as a function of the applied field and temperature, the spin system is initialized from four different configurations: random, skyrmion lattice (SkL), conical (CON), and helical (HL) configurations. The energy is then minimized numerically using stochastic LLG simulation. We calculate the average spin configuration from 2000 spin configurations separated by 10 time steps. The energies of all found configurations are further evaluated and compared with each other to obtain the phase boundaries. Additionally, to avoid boundary-size effects, we checked the calculations for slight variations in the lateral size of the sample in cases close to phase boundaries.

Fig.~\ref{fig1}~(a) shows the magnetic phase diagrams obtained in the simulations for the different film thicknesses, where the HL, CON, and SkL phases are mapped out. It is noteworthy that decreasing the film thickness favors SkL formation over the CON phase, attributed to the increasing significance of surface effects. As our aim is to characterize the skyrmion nucleation from the conical phase, these results will guide us in selecting the appropriate values for magnetic field and temperature to perform further simulations.

\subsection{Nucleation mechanism and activation energies}\label{secIIIB}

To characterize the nucleation and annihilation dynamics of skyrmions during their transition from the conical phase, it is essential to identify both the nucleation mechanism and the energy barriers associated with such a phase transition. Therefore, in this section, we employ the GNEB method (see Sec.\ref{GNEB}) to obtain the corresponding MEPs and activation energies of the mentioned transitions at zero temperature. For the simulations, we consider a sample of size $5L_D\times 5L_D\times d$, initially in the CON phase, and the nucleation of a single skyrmion at the center of the sample. In general, the transition between the two phases can be mediated by different intermediate states. Here, we have considered the following: (i) the formation of chiral bobbers (CB)~\cite{Kim2020mechanism,32,34}, and (ii) the formation of torons~\cite{32}. Fig.~\ref{fig1}~(b) and (c) show the considered intermediate states. The MEPs obtained in our calculations for the case of CB formation in a film of thickness $d=2L_D$ and for different values of the applied field are shown in Fig.~\ref{fig1}~(d), where the reaction coordinate, $x$, defines the normalized (geodesic) displacement along the formation path. Notice that the activation energy for CB nucleation, i.e., the energy difference between the highest energy state (saddle point) and the initial state (CON state), decreases as the applied field decreases and the CB state becomes energetically favorable. This behavior is typical of bistable systems, where two phases coexist as local free energy minima over a particular range of the external field~\cite{bertotti1998hysteresis}. As it will be discussed in more detail later in this section, similar behavior is obtained for different film thickness, i.e. $d=L_D$ and $d=3L_D$. 

\begin{figure*}[!t]
    \centering
    \includegraphics[width=\linewidth]{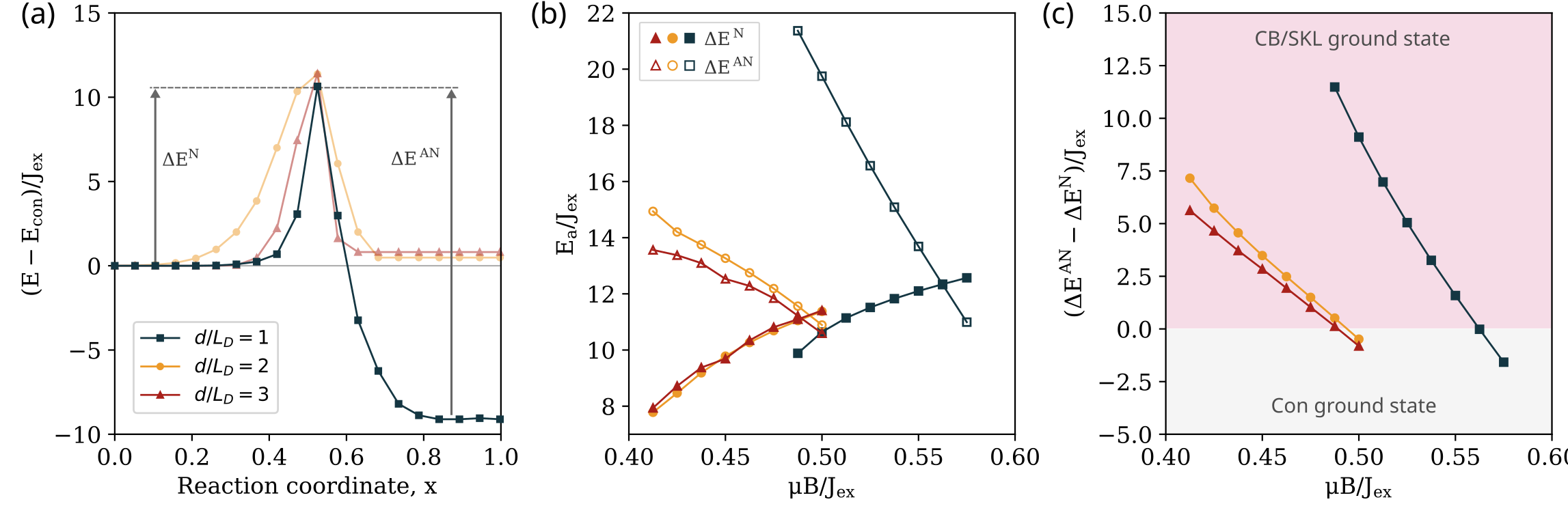}
    \caption{(a) Minimum energy paths for CB nucleation mechanism for different film thicknesses, for applied field $\mu B/J_\text{ex}=0.5$. Gray arrows indicate the activation energies for the CB nucleation ($\Delta E^N$) and annihilation ($\Delta E^{AN}$) for the case of $d=L_D$. (b) Activation energies for CB nucleation (solid symbols) and annihilation (open symbols) as a function of the applied magnetic field, for the three film thicknesses considered in (a). (c) Activation energy difference for the CB nucleation and annihilation, for the three film thicknesses considered in (a). Gray (red) shaded region indicates the parameter space where the conical (CB/SKL) phase is the ground state.}
    \label{fig2}
\end{figure*}

In the case of toron nucleation, the toron phase was found to be unstable for $d=L_D$. On the other hand, for the larger film thicknesses $d=2L_D$ and $3L_D$, although the MEPs can be stabilized, the formation of torons is shown to be energetically unfavorable with respect to the CB formation, as shown by the high activation energies in
the MEPs in Fig.~\ref{fig1}~(e), for $d=2L_D$. As discussed in Ref.~\citenum{32}, the surface effects play an important role in the stabilization of skyrmion tubes in a conical background. Particularly, the formation of torons is favored when increasing the ratio $d/L_D$. Therefore, for magnetic films of thickness $d\leq3L_D$, as considered in this work, we expect the preferential intermediate state for skyrmion nucleation [directly observed in Ref.~\citenum{Kim2020mechanism} with about $60\%$ of the contrast of a fully formed skyrmion] to correspond to CBs instead of torons. From here on, we therefore focus on the formation of CBs as the predominant mechanism for skyrmion nucleation in chiral magnetic thin films.    

Fig.~\ref{fig2}~(a) shows the calculated MEP for CB nucleation for different film thicknesses, for applied field $\mu B/J_\text{ex}=0.5$. The activation energies for the nucleation ($\Delta E^N$) and annihilation ($\Delta E^{AN}$) are marked by gray arrows for the case of $d=L_D$. Notice that, by decreasing the film thickness, the activation energy for CB nucleation suffers minor changes as the saddle point of the phase transition corresponds to the formation of a Bloch point~\cite{rybakov2015new} located only at the film surface, without major bulk contribution. The energy of the fully formed CB state, on the other hand, is strongly affected as the film thickness becomes of the order of the CB size ($\approx L_D/2$)~\cite{32}, as shown for the $d=L_D$ case. In this way, the activation energy for CB annihilation can be strongly affected. 

Fig.~\ref{fig2}~(b) displays the calculated activation energies for CB nucleation (solid symbols) and annihilation (open symbols) as a function of the applied magnetic field, considering the three specified film thicknesses. The activation energies exhibit significant sensitivity to variations in the magnetic field. For instance, reducing the field by $0.075J_\text{ex}$ from the point where $\Delta E^{N}=\Delta E^{AN}$ (where CB and CON states are energetically degenerate) results in an approximate $20\%$ decrease in the activation energies for CB nucleation across all thicknesses. As previously mentioned, $\Delta E^{AN}$ is notably influenced at reduced thicknesses, with the same change in field inducing an increase of approximately $70\%$ in $\Delta E^{AN}$ for $d=L_D$, and around $20\%$ for $d=2L_D$ and $3L_D$.

In all scenarios, the difference between nucleation and annihilation energies follows a nearly linear relation as the magnetic field approaches the phase boundary, as shown in Fig.~\ref{fig2}~(c). Furthermore, it is noticeable that increasing the film thickness from $d=2L_D$ to $3L_D$ results in small alterations in the activation energies, as the CB state predominantly resides near the film surface. Consequently, although the energy of the fully formed SKL state is significantly influenced by the value of $d$, one might anticipate that further increases in film thickness will yield comparable CB activation energies to those observed for $d=3L_D$.

\begin{figure*}[!t]
    \centering
    \includegraphics[width=\linewidth]{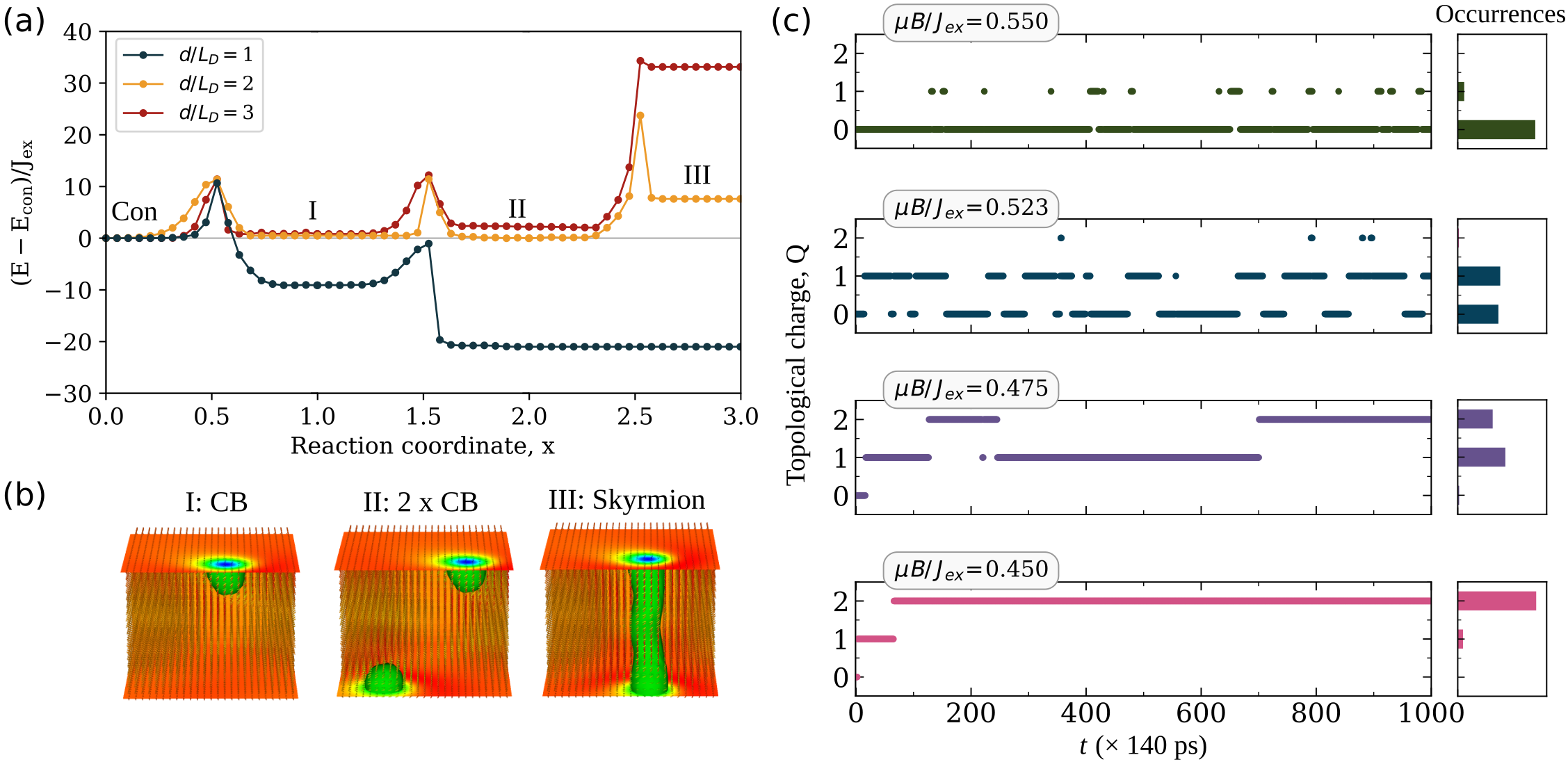}
    \caption{(a) Minimum energy path for the nucleation of a single skyrmion from the conical phase, for different film thicknesses and for applied field $\mu B/J_\text{ex}=0.5$. The first and second peaks along the formation paths correspond to the nucleation of CBs at the top and bottom surfaces. For $d>L_D$, a third peak along the formation path corresponds to the connection of the CBs into the skyrmion tube. (b) The stable states along the MEPs in (a). (c) Topological charge $Q$ calculated at the film surface as a function of the simulation time, for different values of applied field, for thickness $d=2L_D$ and temperature $k_B T/J_\text{ex}=0.7$. Histograms on the right-hand side show the normalized number of occurrences of the states with $Q=0$ (CON phase), $Q=1$, and $Q=2$.}
    \label{fig3}
\end{figure*}

\subsection{From chiral bobber to skyrmion}\label{sec.IIIC}

Once in the CB state, the formation of the skyrmion can proceed to the next stages. Fig.~\ref{fig3}~(a) shows the minimum energy paths obtained in our calculations for the nucleation of a single skyrmion from the conical phase, for different film thicknesses, at a fixed magnetic field $\mu B/J_{ex}=0.5$. The first and second peaks along the formation paths correspond to the nucleation of the Bloch points of CBs at the top and bottom surfaces. For $d>L_D$, a third peak along the formation path corresponds to the connection of the CBs into the skyrmion tube. Since the CBs have a typical depth of $L_D/2$, the nucleation of CBs at both surfaces in the case of $d=L_D$ is sufficient to connect the skyrmion tube across the sample; therefore, for that case, only two peaks are seen along the MEP. Fig.~\ref{fig3}~(b) shows the stable states encountered along the formation path.

Notice that, as discussed in the previous section, the activation energies for the nucleation of CBs are not sensitive to the film thickness. However, the energy cost for the formation of the skyrmion tube strongly increases with $d$, as the skyrmion energy is proportional to its length. Therefore, one can observe that, once in the CB state, the energy barrier for the system to transition to the skyrmion phase can be much larger than the barrier to return to the conical phase, as illustrated, for instance, in the MEPs for $d=2L_D$ and $3L_D$ in Fig.~\ref{fig3}~(a). This can lead to sequential instances of nucleation and collapse (blinking behavior) of CBs during the formation of the skyrmion lattice. As discussed in Sec.~\ref{secIIIB}, the activation energies can be tuned up and down by the applied magnetic field, in a way that one can seek the optimal parameters that support the blinking behavior. As we will show in the next section, the blinking is favored when the system is close to, but not exactly at, the phase boundary between the SkL and CON states.

\section{Skyrmion blinking}\label{sec.IV}

\subsection{Spin dynamics simulations}\label{sec.IVA}

To characterize the dynamics of the blinking process in more detail, we employ spin dynamics simulations based on stochastic LLG approach (as detailed in Sec.~\ref{secIIA}). To monitor the nucleations and collapses of CBs over time in our simulations, we compute the two-dimensional topological charge, defined as $Q=\frac{1}{4\pi}\int\textbf{n}\cdot\left(\frac{\partial\textbf{n}}{\partial x}\times \frac{\partial\textbf{n}}{\partial y} \right)dxdy$, at the film surface. This enables easy differentiation between the CON phase, where $Q=0$, and skyrmionic states, where each CB or skyrmion carries a topological charge of $Q=1$. For simplicity, this section considers a simulation box capable of accommodating a maximum of two skyrmions only, with the damping parameter set to $\alpha=0.3$. The spin system is initialized in the conical phase, and the skyrmions are allowed to nucleate and occupy the sample. Magnetic configurations are saved every $140$~ps (with simulation time step $\Delta t=10^{-15}$~s), derived from the average of $2000$ configurations obtained during the final $40$~ps of each period. Fig.~\ref{fig3}~(c) shows the topological charge calculated at the film surface as a function of simulation time for various values of the applied field, for temperature $k_B T/J_\text{ex}=0.7$ and film thickness $d=2L_D$. Notice that, for the highest field considered in Fig.~\ref{fig3}~(c) ($\mu B/J_{ex}=0.55$), the CON phase ($Q=0$) is the lowest energy state, and CBs rarely nucleate and are rapidly destroyed. Consequently, the system predominantly resides in the CON state. Conversely, for the lowest field ($\mu B/J_{ex}=0.45$), CBs rapidly nucleate, and the system quickly transitions to the SkL phase, with a stable topological charge of $Q=2$. At intermediate fields, the system encounters scenarios where the energy barriers for transitioning a CB phase to the skyrmion phase can be significantly larger than the barrier to return to the conical phase, resulting in a sequence of CB nucleations and collapses (blinking behavior). Specifically, for $\mu B/J_{ex}=0.523$, we observed the system spending $50\%$ of the time in the CON phase and the other $50\%$ in the CB phase, as depicted by the histograms in Fig.~\ref{fig3}~(c). This condition characterizes uniform blinking behavior, where neither the CON phase nor the CB/skyrmion phase predominates. Interestingly, this condition is achieved for a field value significantly above the critical field, $B_c$, where the energy barriers for nucleation and annihilation become equal ($\mu B_c=0.4941$~$J_{ex}$ for $d=2L_D$). As we will discuss further in this section, this feature stems from the different attempt frequencies associated with the CB nucleation and annihilation processes.

\begin{figure}[!t]
    \centering
    \includegraphics[width=\linewidth]{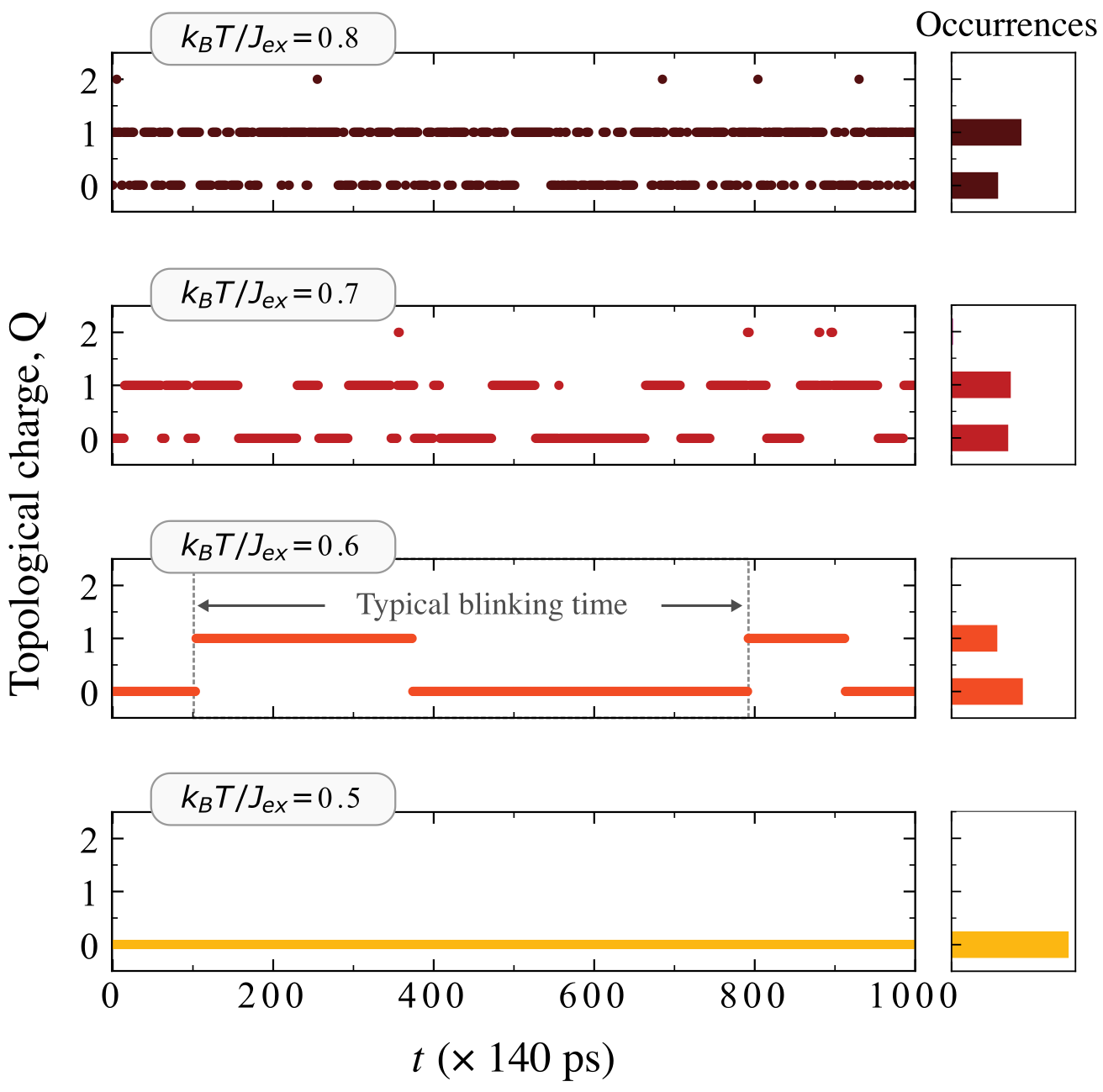}
    \caption{Topological charge $Q$ calculated at the film surface as a function of the simulation time, for different temperatures, for film thickness $d=2L_D$ and applied field $\mu B/J_\text{ex}=0.523$. Histograms on the right-hand side show the normalized number of occurrences of the states with $Q=0$ (CON phase), $Q=1$, and $Q=2$.}
    \label{fig4}
\end{figure}

\begin{table}[b!]
\begin{tabular}{|c|c|c|c|}
\hline
$k_B T/J_\text{ex}$  &$\left<\tau\right>^\text{N}$ (ns)  &$\left<\tau\right>^\text{AN}$ (ns)& $\left<\tau\right>^\text{blink}$ (ns) \\ \hline
0.60 & $60\pm8$ & $34\pm6$ & $92\pm9$\\
0.65 & $11\pm2$& $16\pm3$ & $26\pm4$\\
0.70 & $4.7\pm0.7$ & $6.0\pm0.9$ & $11\pm1$\\
0.75 & $1.5\pm0.1$ & $2.2\pm0.2$ & $3.7\pm0.2$\\
0.80 & $0.85\pm0.08$ & $1.3\pm0.1$ & $2.1\pm0.2$\\
\hline
\end{tabular}

\caption{The average transition times for the CB nucleation ($\left<\tau\right>^\text{N}$), annihilation ($\left<\tau\right>^\text{AN}$), and blinking ($\left<\tau\right>^\text{blink}$) processes, calculated over $8000$ intervals of $140$ ps in the simulations, at different temperatures, for magnetic field $\mu B/J_{ex}=0.523$ and film thickness $d=2L_D$.}  
\label{table_parameters}
\end{table}

The condition for uniform blinking is therefore primarily determined by the applied field. On the other hand, the temperature regulates how frequently the transitions occur, thereby influencing the blinking time. Fig.~\ref{fig4} shows the topological charge calculated at the film surface as a function of simulation time at different temperatures, for $\mu B/J_{ex}=0.523$ and $d=2L_D$. A typical blinking time is depicted in Fig.~\ref{fig4} for the scenario where $k_B T/J_\text{ex}=0.6$, representing the time interval between consecutive nucleations of the CB. It is noteworthy that, except for the case of $k_B T/J_\text{ex}=0.5$, where no CBs nucleate during the simulated time window, the system remains divided between the CON and CB phases, with a slight increase of CB occurrences ($Q=1$) with temperature. However, the blinking time significantly decreases with increasing temperature. Table~\ref{table_parameters} summarizes the average transition times $\left<\tau\right>$ obtained in the simulations, at different temperatures, for the CB nucleation, annihilation, and blinking processes, in the case considered.

\begin{figure}[!t]
    \centering
    \includegraphics[width=\linewidth]{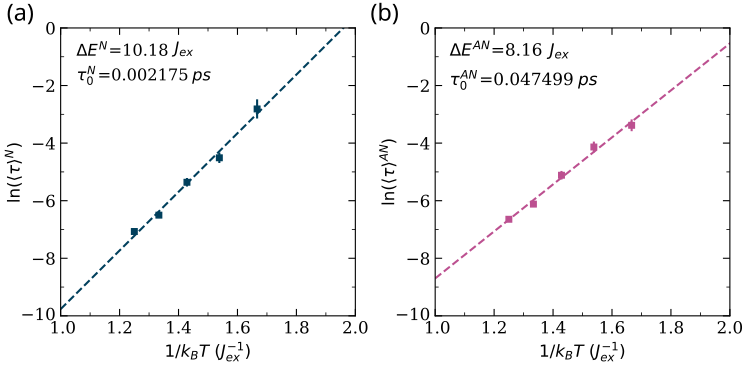}
    \caption{The dependence of $\ln(\left<\tau\right>)$ on $1/k_BT$ for (a) nucleation and (b) annihilation processes of chiral bobbers, for values taken from Table~\ref{table_parameters}. The values of $\tau_0$ and $\Delta E$ indicated in (a) and (b) are obtained by linear fitting of the data, as shown by the dashed lines.}
    \label{fig5}
\end{figure}
\begin{figure*}[!t]
    \centering
    \includegraphics[width=0.85\linewidth]{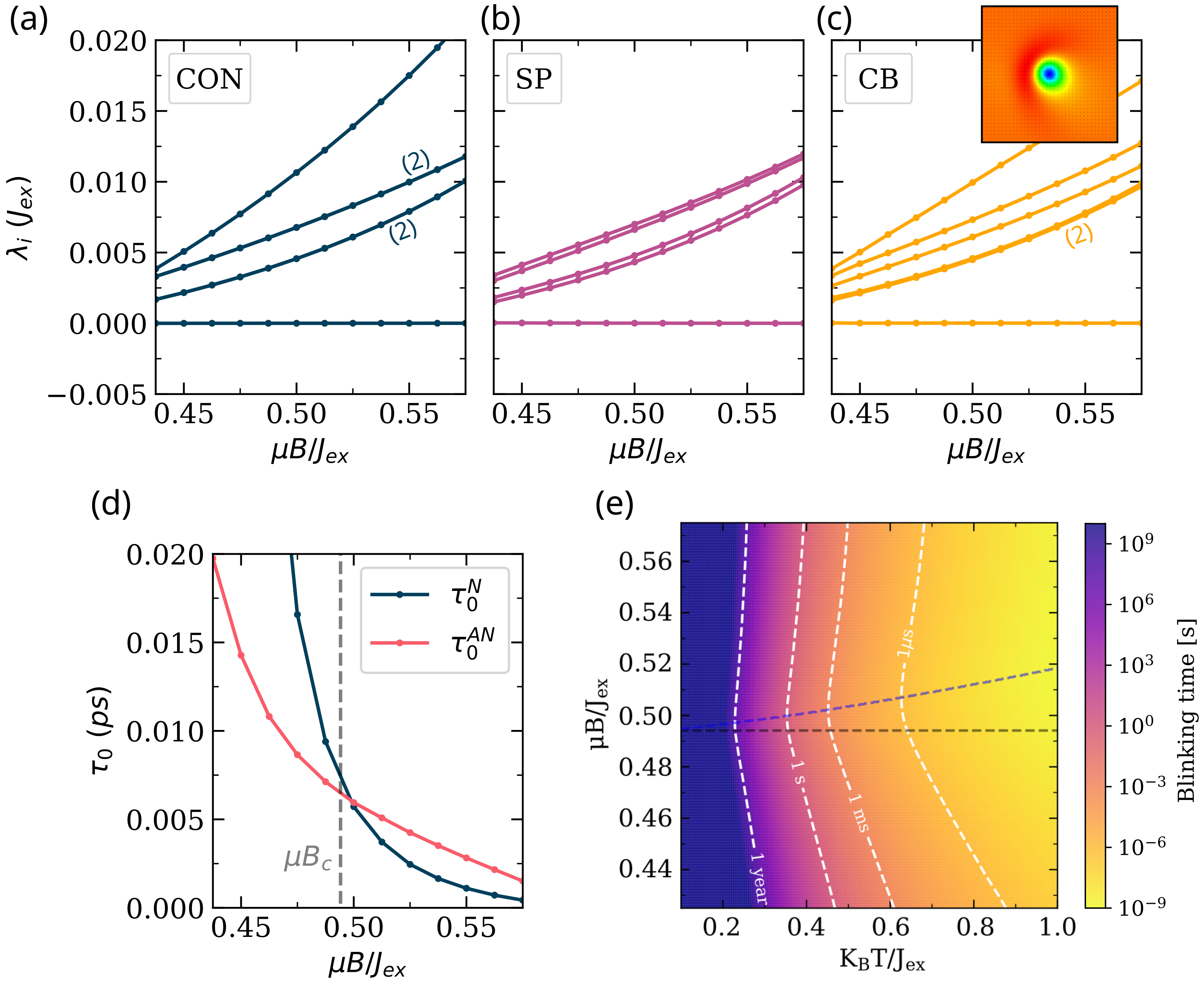}
    \caption{(a-c) Eigenvalue spectra obtained as a function of the applied magnetic field, for the CON, SP and CB states. For better visualization, the negative eigenvalue of the SP state is not shown. Numbers in brackets indicate the number of degenerate (or quasi-degenerate) modes. The inset in (c) illustrates the $z$-component of the magnetization in a cross section of the CB state. (d) Pre-exponential factor calculated from HTST, for both CB nucleation ($\tau_0^N$) and annihilation ($\tau_0^{AN}$) processes, as a function of the magnetic field. (e) Blinking time, $\tau^\text{blink}=\tau^N+\tau^{AN}$, obtained from HTST as a function of applied field and temperature. The black dashed line indicates the critical field, $B_c$, where the energy barriers for nucleation and annihilation become equal. The blue dashed line indicates the uniform blinking behavior.}
    \label{fig6}
\end{figure*}

The average transition times obtained in the simulations can be related to the phase transition rates predicted by HTST (see Eq.~\eqref{eq.HTST}), where one can write $\tau=1/\Gamma=\tau_0\exp(\Delta E/k_BT)$, with $\tau_0=1/\Gamma_0$. As depicted in Fig.~\ref{fig5}, we observe a linear behavior of $\ln(\left<\tau\right>)$ with $1/k_BT$, for both nucleation and annihilation processes, which indicates a temperature-independent pre-exponential factor $\Gamma_0$. According to Eqs. \eqref{eq.prefactor1} and \eqref{eq.prefactor2}, that corresponds to a system with the same number of zero modes at the minimum and saddle point states. 

Moreover, by linear fitting of the data, we can estimate the effective energy barriers and pre-exponential factors associated with the CB nucleation and annihilation processes in our spin dynamics simulations. The obtained energy barriers, indicated in Fig.~\ref{fig5}~(a) and (b), are similar to those obtained by the GNEB method for the same value of applied field ($\Delta E^N_{GNEB}=11.94J_{ex}$ and $\Delta E^{AN}_{GNEB}=9.38J_{ex}$), and a higher pre-exponential factor is observed for the CB annihilation than for the nucleation. The latter justifies the observation of uniform blinking for $B>B_c$, as equal activation energies can result in different transition times due to the prefactors. One should notice that the transition rates obtained by LLG dynamics may differ for different damping parameter $\alpha$~\cite{bessarab2012harmonic}, or in presence of boundary effects, among other conditions that can influence the system differently from the nucleation of an isolated CB.

It is noteworthy that the collapse of CBs differs from that of skyrmions in thin films, where the pre-exponential factor linearly depends on temperature~\cite{muller2019spirit, desplat2018thermal, von2019skyrmion}. In the case of two-dimensional skyrmions, the presence of two translational zero modes contrasts with the absence of such zero modes in the defect-like saddle point, resulting in a temperature dependent pre-exponential factor, as predicted in Eq.~\eqref{eq.prefactor2}~\cite{bessarab2018lifetime, buijnsters2014zero}. On the other hand, for CBs, the system also possesses a defect-like Bloch point. This, combined with lattice discretization, results in nonzero translational modes. As we will discuss below, another zero mode, distinct from translational ones, exists in the CB state.

\subsection{Transition rates from HTST}\label{sec.IVB}

To determine the transition rates of the CB within the HTST framework, we compute the eigenvalues $\lambda_i$ of the Hessian matrix corresponding to the local minima and saddle points along its formation and annihilation MEPs. Figure~\ref{fig6} (a-c) presents the six lowest energy eigenvalues obtained from our calculations as a function of the applied field. In both the minima and saddle point states, only one zero mode is observed. This zero mode arises due to the system's degeneracy through rotations of the magnetization perpendicular to the direction of the applied field, corresponding to the angular phase of the conical state.

When a skyrmion or CB is embedded in the conical state, its domain-wall region adjusts to the conical phase, resulting in asymmetry across the cross section perpendicular to the applied field, as shown, for instance, in the inset of Fig.~\ref{fig6}~(c) and discussed in previous works~\cite{leonov2017asymmetric,leonov2023mechanism}. Consequently, the zero mode also characterizes the rotation of CB and saddle point states. 

This result is consistent with our spin dynamics simulations, where a temperature-independent pre-exponential factor was determined, indicating a system with an equal number of zero modes at both the minimum and saddle point states. Figure~\ref{fig6}~(d) illustrates the pre-exponential factor $\tau_0=1/\Gamma_0$, computed from Eqs.~\eqref{eq.prefactor1} and \eqref{eq.prefactor2}, along with the corresponding eigenvalues as a function of the applied field, for both CB nucleation and annihilation processes. Similarly to the collapse of a skyrmion in thin films~\cite{von2019skyrmion}, the pre-exponential factor undergoes significant changes with the magnetic field, thereby playing an important role in determining the rate of the phase transition. It is worth noting that, consistent with the spin-dynamics simulations, the HTST also predicts a higher pre-exponential factor $\tau_0$ for the CB annihilation process compared to nucleation when $\mu B>\mu B_c$. 

By combining the calculated pre-exponential factors with the activation energies obtained via the GNEB method, we can estimate the nucleation and annihilation times using Eq.~\eqref{eq.HTST}. Figure~\ref{fig6}~(e) illustrates the resulting blinking time, $\tau^\text{blink}=\tau^N+\tau^{AN}$, as a function of applied field and temperature.

It is noteworthy that, for each temperature, there exists a field value, denoted as $B^\ast$, at which the blinking time is minimized, as depicted by the blue dashed line in Fig.~\ref{fig6} (e). This condition corresponds to $\tau^N=\tau^{AN}$, signifying uniform blinking behavior, wherein CBs nucleate and collapse at the same rate. Such uniform blinking is achieved at a field value above the critical field, $B_c$ (indicated by the black dashed line in Fig.~\ref{fig6}~(e)), where the energy barriers for nucleation and annihilation become equal. This finding aligns with our spin-dynamics simulations, where, for $\mu B=0.523 J_{ex}>\mu B_c$, we observed the system spending 50\% of the simulated time in the CON phase and the other 50\% in the CB phase (see Fig.~\ref{fig3}~(c)). This observation stems from the distinct pre-exponential factors associated with CB nucleation and annihilation processes.

In essence, our results suggest that by appropriately tuning the applied field and temperature, uniform blinking of metastable chiral bobbers can be achieved. A similar behavior is expected for other film thicknesses, in the vicinity of the phase boundary between CON and SKL states. 

\begin{figure*}[!t]
    \centering
    \includegraphics[width=0.8\linewidth]{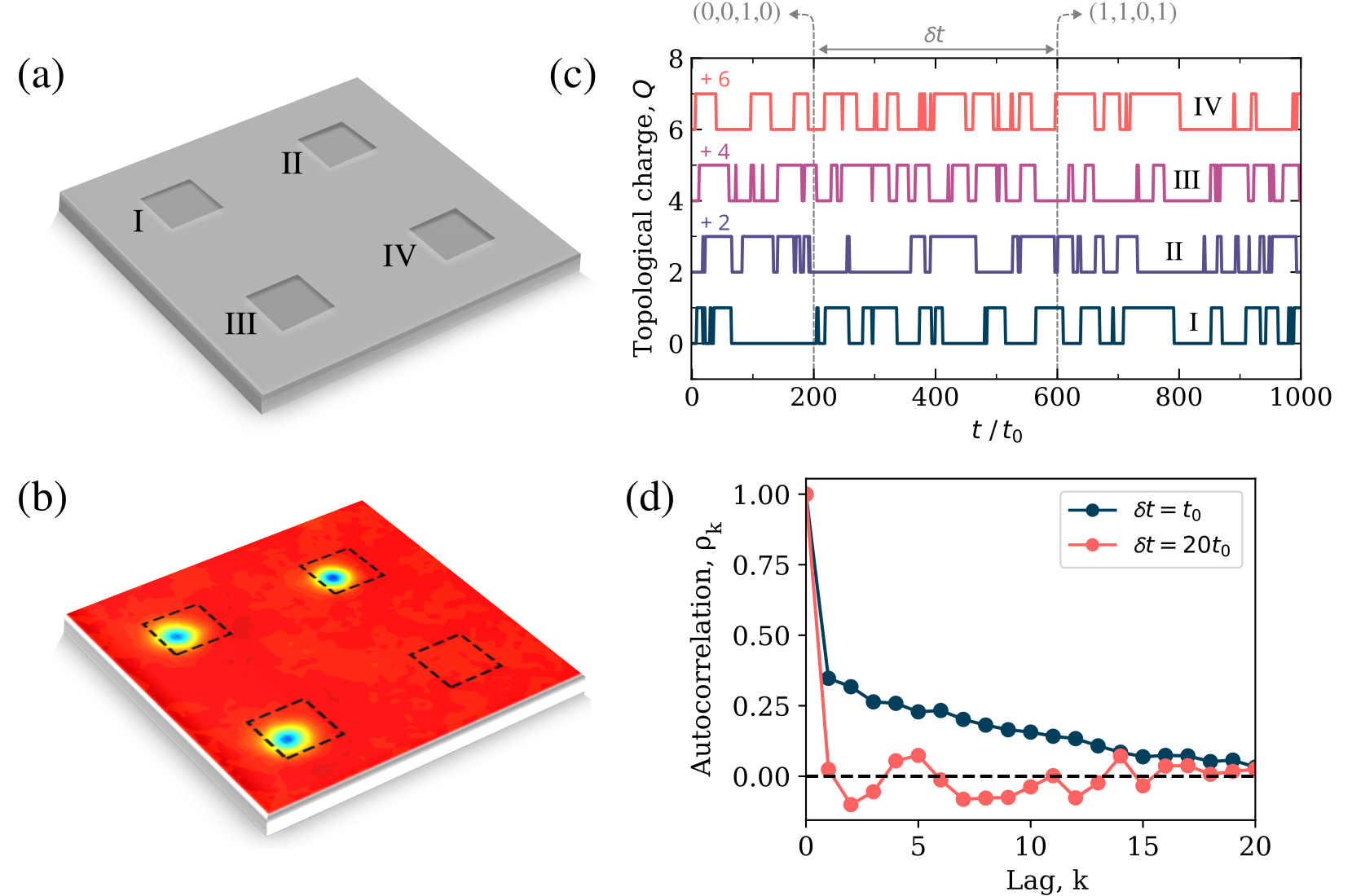}
    \caption{(a) Schematic illustration of a $2\times2$ array of blind holes (labelled I to IV), considered to simulate the random number generator device. (b) For the appropriate applied field and temperature, skyrmions and CBs nucleate solely inside the blind holes. (c) Topological charge $Q$ calculated as a function of time (where $t_0=140$~ps), for each blind hole in (a) separately. The curves are shifted in $Q$ for better visualization. Two examples of combinations ($Q_{I}$, $Q_{II}$, $Q_{III}$, $Q_{IV}$), measured at $t/t_0=200$ and $t/t_0=600$, are indicated in gray, as well as the time frame between consecutive measurements, $\delta t$. (d) The autocorrelation function as a function of the lag, calculated in the simulations for different values of $\delta t$. }
    \label{fig7}
\end{figure*}

\section{Skyrmion blinking for probabilistic
computing}\label{secV}

The switching between skyrmionic and non-topological states is well established as a promising bit operation for information transport and storage~\cite{hagemeister2015stability}. Understanding the temperature, thickness, and field dependence of the nucleation and collapse processes of magnetic skyrmions is therefore crucial to the development of novel technological applications. Of particular recent interest is the thermally driven dynamics of skyrmions, that can be used for probabilistic computing devices, such as signal reshuffling devices~\cite{pinna2018skyrmion,zazvorka2019thermal} and random number generators~\cite{wang2022single}. Both latter applications rely on the ability of the thermally driven skyrmion system to generate uncorrelated copies of itself. In this section, we illustrate the applicability of skyrmion blinking in an alternative design of skyrmion-based random-number generators. Since the stochastic nucleation and collapse of CBs in thin films can be fine-tuned by temperature, film thickness and applied magnetic field, as discussed in the previous sections, skyrmion blinking emerges as a viable base ingredient for controlled probabilistic computing.

To demonstrate such a functionality, we envision the device setup presented in Fig.~\ref{fig7}~(a), where an array of blind holes (regions with reduced thickness) is patterned in the magnetic film. Since the magnetic phase diagram is thickness dependent (see Fig.~\ref{fig1}~(a)), by tuning the magnetic field and temperature, one can achieve the scenario where skyrmions/CBs nucleate only at the regions of the sample with reduced thickness, as illustrated in Fig.~\ref{fig7}~(b), and excite the skyrmion blinking behavior at the pre-selected locations of the sample where blind holes are positioned. Considering the example of a $2\times2$ array of blind holes, as illustrated in Fig.~\ref{fig7}~(a-b), since the skyrmions are located far apart from each other, each skyrmion experiences a stochastic blinking behavior that is uncorrelated with the others. By measuring the skyrmion state at different times, one would obtain, in this example, random sequences of four binary digits, each digit representing the presence or absence of a skyrmion in each blind hole.  

We perform micromagnetic simulations of such a system by considering a magnetic film of size $6L_D\times 6L_D\times 2L_D$, with four blind holes of size $1L_D\times 1L_D\times 1L_D$, centered in the four quadrants of the sample (Fig.~\ref{fig7}~(a)). For the simulations, we consider temperature $k_B T/J_\text{ex}=0.7$ and the applied field $\mu B/J_{\text{ex}}=0.555$, which yields uniform blinking behavior for the case of film thickness $d=L_D$ (as inside the blind holes), but does not favor the formation of skyrmions/CBs in the regions where $d=2L_D$ (outside the blind holes). For comparison, the considered field value here is higher than the highest field in Fig.~\ref{fig3}~(c), where CBs rarely nucleate for $d=2L_D$ under the same temperature. 
Fig.~\ref{fig7}~(c) shows the calculated topological charge $Q$ as a function of time for each blind hole separately. Note that the topological charge of each blind hole (labelled I to IV) independently oscillates as the skyrmions undergo a blinking behavior. By measuring the topological charges at different times, one can therefore obtain random sequences of four binary digits, from a total of 16 possible combinations. This demonstrates the basic functionality of the system as a random number generator. Two examples of sequences ($Q_{I}$, $Q_{II}$, $Q_{III}$, $Q_{IV}$) are indicated in Fig.~\ref{fig7}~(c), measured at simulation times $t/t_0=200$ and $t/t_0=600$, with $t_0=140$~ps.

Importantly, as the skyrmions have a characteristic blinking time, to ensure that the obtained random numbers are uncorrelated, one should ensure that the time frame between consecutive measurements, defined here as $\delta t$, is not much smaller than the typical blinking time $\tau^\text{blink}$. To verify this, we calculate the auto-correlation function (ACF) of a sequence of $N$ measured outputs, representing a time-dependent, discrete random variable $X_t=\{x_1, x_2, x_3, ..., x_N\}$, where $x_i\in\{1,2,...,16\}$ corresponds to one of the 16 possible output states, obtained at the $i^{th}$ measurement. The ACF describes the correlation between observations of the time series at two points in time, separated by a specific lag, denoted here as $k$. Essentially, it quantifies how a value in $X_t$ is related to its previous values. The ACF is defined as the ratio of the covariance of $X_t$ and $X_{t+k}=\{x_{k+1}, x_{k+2}, ..., x_N\}$, and the variance of $X_t$, and can be written as~\cite{priestley1981spectral}

\begin{equation}
    \rho(k) =\frac{N}{N-k}\frac{\sum_{i=1}^{N-k}\left(x_i-\Bar{X_t} \right)\left(x_{i+k}-\Bar{X_t}\right)}{\sum_{i=1}^{N}\left(x_i-\Bar{X_t} \right)^2},
\end{equation}
where $\Bar{X_t}$ is the mean value of $X_t$. An autocorrelation $\rho(k)=0$ means that the measured states ${x_i}$ can be considered independent of each other, or uncorrelated. Fig.~\ref{fig7}~(d) shows the ACF as a function of the lag $k$, calculated in the simulations, in two scenarios: (\textit{i}) when the time frame between consecutive measurements is given by $\delta t=t_0$, and (\textit{ii}) when $\delta t=20t_0$. Notice that, in the first scenario, since $\delta t\ll\tau^\text{blink}$ (in the simulations, $\langle\tau\rangle^\text{blink}\approx 40 t_0$), consecutive measurements of the skyrmion state are likely to yield the same output configuration, as the system did not have sufficient time to experience nucleations and collapses. This results in high values for the ACF, as shown in Fig.~\ref{fig7}~(d). Such correlation decreases as one increases the lag, approaching $\rho(k)=0$ for $k>20$, where the time shift between the compared states becomes large enough for the system to experience oscillations. In the second scenario, where $\delta t=20t_0$ ($\approx 50\%$ of the blinking time), we obtain $\rho(k)\approx0$ for every $k>0$, thus demonstrating that one can obtain a sequence of uncorrelated random numbers with the proposed setup, as long as the time frame between consecutive measurements is larger than $50\%$ of the typical blinking time.     

Furthermore, since the number of possible output configurations scales exponentially with the number of blind holes (i.e., $2^n$ possible states for $n$ blind holes), similar setups could be envisioned for generating a wide range of uncorrelated random numbers. For instance, a $4\times4$ array of blind holes would be able to generate $65536$ different states. 


\section{CONCLUSION}
\label{Sec:Conclusion}

In summary, we have comprehensively characterized the dynamic process of skyrmion nucleation from the conical phase in helimagnets. Using a geodesic nudged elastic band (GNEB) method and spin dynamics simulations, we investigated different nucleation mechanisms and calculated the activation energies for skyrmion formation as a function of film thickness and applied magnetic field. We revealed that a blinking behavior (a creation-destruction process) of skyrmions is favored by the local stability of chiral bobber (CB) states. We demonstrated that such states can be set to uniform blinking when excited by the appropriate applied field and temperature. This condition is particularly favored near the phase boundary between the skyrmion lattice and conical states, specifically on the conical side of the phase boundary, where CBs are metastable.
We calculated the oscillation modes of such CB states and demonstrated that their collapse fundamentally differs from that of skyrmions in thin films, as only one rotational zero mode is present, in contrast to the two translational zero modes of skyrmions.
Combining harmonic transition state theory and GNEB methods, we estimated the typical blinking frequencies of CBs (ranging from KHz to GHz) as a function of the applied field and temperature.
Finally, we exemplified the use of skyrmion blinking for controlled probabilistic computing and demonstrated the basic functionalities of a skyrmion-based random-number generator. Together, our results advance the understanding of the nucleation mechanisms of skyrmions in chiral magnets and, given the expected reproducibility of our findings in readily existing experimental setups, we anticipate that this work will spur further experimental investigations into skyrmion dynamics and further development of skyrmion-based applications.

\section*{Acknowledgements}
This work was supported by the Research Foundation - Flanders (FWO-Vlaanderen) and Brazilian agencies FACEPE, CAPES and CNPq.

\bibliographystyle{unsrt}
\bibliography{references.bib}


\end{document}